\documentclass[doublecol]{epl2} 

\newcommand{\be}{\begin{equation}}
\newcommand{\ee}{\end{equation}}
\newcommand{\bs}{\begin{subequations}}
\newcommand{\es}{\end{subequations}}

\usepackage{color}

\title{The influence of Lifshitz forces and gas on premelting of ice within porous materials}

\author{M. Bostr{\"o}m \inst{1,2,a} \and O. I. Malyi \inst{3,a} \and  P. Thiyam \inst{4,a} \and  K. Berland \inst{1}  \and I. Brevik \inst{2} \and C. Persson \inst{1,4,5}  \and D. F. Parsons \inst{6}  }

\institute{ 

\inst{1}Centre for Materials Science and Nanotechnology,
University of Oslo, P. O. Box 1048 Blindern, NO-0316 Oslo, Norway\\
\inst{2} Department of Energy and Process Engineering, Norwegian University of Science and Technology, NO-7491 Trondheim, Norway\\
\inst{3} School of Materials Science and Engineering, Nanyang Technological University, 50 Nanyang Avenue, Singapore 639798, Singapore\\
 \inst{4} Department of Materials Science and Engineering,
Royal Institute of Technology, SE-100 44 Stockholm, Sweden\\
\inst{5}Department of Physics, University of Oslo,
P. O. Box 1048 Blindern, NO-0316 Oslo, Norway\\
\inst{6}School of Engineering and IT, Murdoch University,
90 South St, Murdoch, WA 6150, Australia\\
\inst{a} Email corresponding authors: Mathias.Bostrom@smn.uio.no; oleksandrmalyi@gmail.com; thiyam@kth.se\\}

 \shortauthor{M. Bostr{\"o}m  \etal}


\abstract{Premelting of ice within pores in earth materials is shown to depend on the presence of vapor layers.
For  thick vapor layers between ice and pore surfaces, a nanosized water sheet can be formed due to repulsive Lifshitz forces. 
In the absence of vapor layers, ice is inhibited from melting near pore surfaces. In
between these limits, we find an enhancement of the water film thickness in silica and alumina pores. 
In the presence of metallic surface patches in the pore, the Lifshitz forces can dramatically widen the water film thickness, with potential complete melting of the ice surface.
 }

\begin{document}

\maketitle

\begin{figure}
\includegraphics[width=8cm]{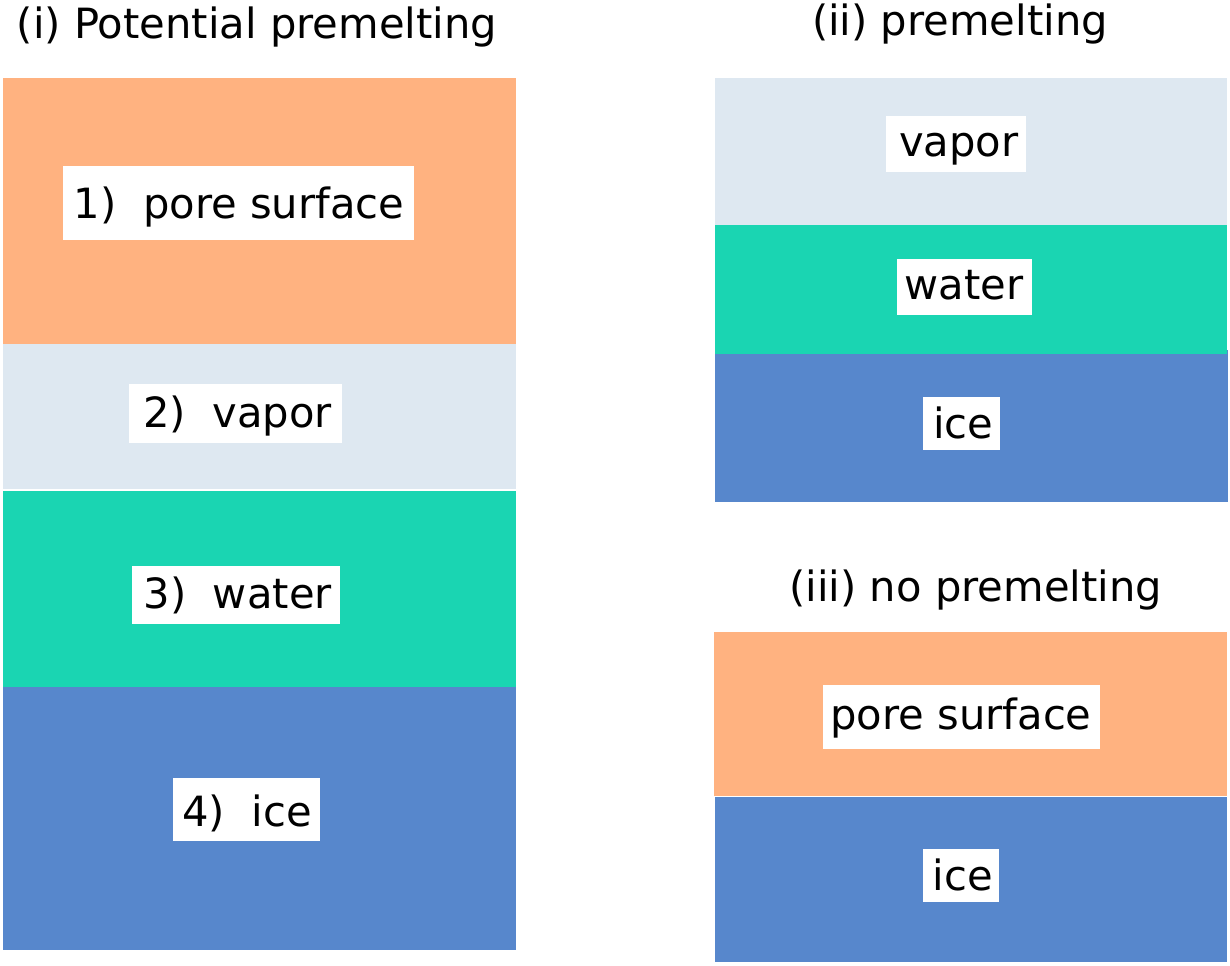}
\caption{(Color online)  (i) Model system where transition between repulsive and attractive Lifshitz interaction can cause ice to premelt in a gas filled pore within rock or clay.
 (ii)  The role of gas is to go towards a system with ice-water-gas where there will be a repulsive-attractive transition which gives a finite size premelted water film as predicted by Elbaum and Schick.\,\cite{Elbaum}  (iii)  In the absence of gas the system with ice-water-pore will have short range attraction which leads to a reduction of water film towards zero film thickness. In a pore with metallic surface patches, repulsive contributions for metal-gas-water-ice can lead to complete surface melting for a range of nanometer sized vapor layer thicknesses. }
\label{figu1}
\end{figure}

At the interface between ice and a dissimilar material, interfacial melting can occur. In some cases this melting can become unbounded resulting in complete interfacial melting of ice. This case is here referred to as premelting in the case of the vapor-ice interface. 
\cite{Elbaum,Wilen}  In porous media, premelting of ice is here shown to depend on the presence of diluted gas layers between ice and pore surface as well as on
the optical properties of the surface material.  This ought to be  important to account for in any realistic modeling that aims at understanding how liquid  water can be present in frozen pores on icy planets and in permafrost regions. 
Different surface materials including silica (serving as  model for rock composed of quartz), alumina, and gold are studied. 
We show how enhanced concentrations of  liquid water can be induced in pores depending on the nature of the surface materials and the thickness of the vapor layers.
Our model system is shown in Fig.\,\ref{figu1}. 
It is well established that  Lifshitz forces acting between surfaces can be either attractive or repulsive depending on the material composition and geometry.\,\cite{ Lond, Casi, Maha,Ser, Milt, Pars, Ninhb,bosserPRA2012,Esteso}
To enable accurate calculations of Lifshitz energies, a detailed knowledge of the dielectric functions is required.\,\cite{Maha,Zwol1,Mund,SashaPCCP2016,bosserPRA2012,Esteso,Elbaum} 
In a subtle experiment, Hauxwell and Ottewill\,\cite{Haux} measured the thickness of oil films on water near the alkane saturated vapor pressure, which depends on Lifshitz forces.\,\cite{Rich, Haux, Ninh} 
Experiments studying liquid helium films on planar surfaces \,\cite{AndSab} and later theoretical analysis have also demonstrated  the validity of the Lifshitz theory.\,\cite{Maha,Rich1,Haux,Ninh}

Careful measurements of repulsive Lifshitz forces have been reported in the literature.\,\cite{Mund,Zwol1,Milling,Lee,Feiler}
Lifshitz energy is calculated as a sum of frequency contributions. 
Frequency intervals where the intervening medium has a dielectric permittivity in between the permittivity of the two interacting objects give rise to a repulsive contribution; other intervals give rise to an attractive contribution.\,\cite{Maha} 
As the properties of the objects vary, such as the gas film thickness, the attractive and repulsive frequency contributions
may cancel, causing a transition from an attractive
to repulsive Lifshitz force.
In the process  of ice melting,  at the triple point a nanosized  film of water  has been predicted to be stabilized by repulsive Lifshitz forces.\,\cite{Elbaum,Wilen,Elbaum1993,Mohl,Dash} For other conditions (e.g. other temperatures or presence of salt ions\,\cite{Wettlaufer,BWN2001,Zhang,Thiyam}), thicker or thinner films could occur. 
It  has  in the past been found that by coating the ice with a thin hydrocarbon layer one can substantially increase the water sheet thickness on the surface of melting ice.\,\cite{Bar-Ziv} 
Here we rely on calculations on Lifshitz forces
to study the problem of premelting of ice within gas-filled frozen porous media.
For pores with metallic surface patches, our results suggest that in the absence of a vapor layer, the ice will stay dry.
For thick vapor films, the results approach  those of ice in free space with a film of liquid water
at the ice surface. 
However, for a certain range of thicknesses of the vapor layer situated between the ice and metal surface,  complete surface melting of ice can occur. 
Our results illustrate that model calculations can provide insight into how gas molecules, i.e. vapor layers near pore surface, can induce the formation of liquid water within frozen pores.

\begin{figure}
\includegraphics[width=8cm]{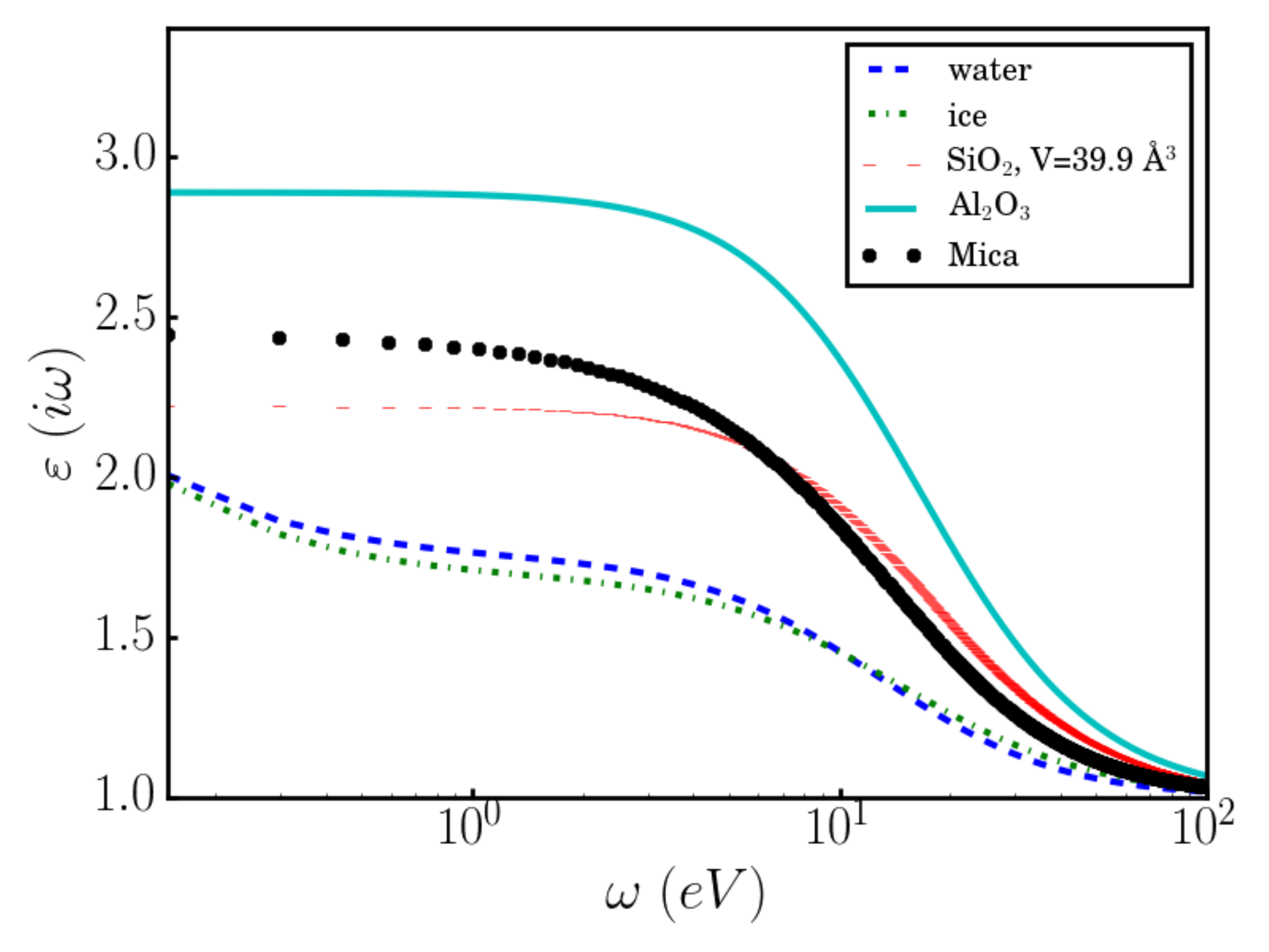}
\caption{ (Color online) The dielectric functions at imaginary frequencies are shown for different materials.   The static values, $\epsilon(0)$, are 91.5, 88.2, 4.1,  10.3, and 2.45 for ice, water, quartz, alumina, and mica.  We use SiO$_2$ with $V$=~39.9 \AA$^3$ as a model for quartz where $V$ is the volume per SiO$_2$ unit as defined by Malyi et al.\,\cite{SashaPCCP2016}. The crossing of the curves for ice and water is essential for retardation driven repulsive-attractive pressure transitions}
\label{figu2a}
\end{figure}

\begin{figure}
\includegraphics[width=8cm]{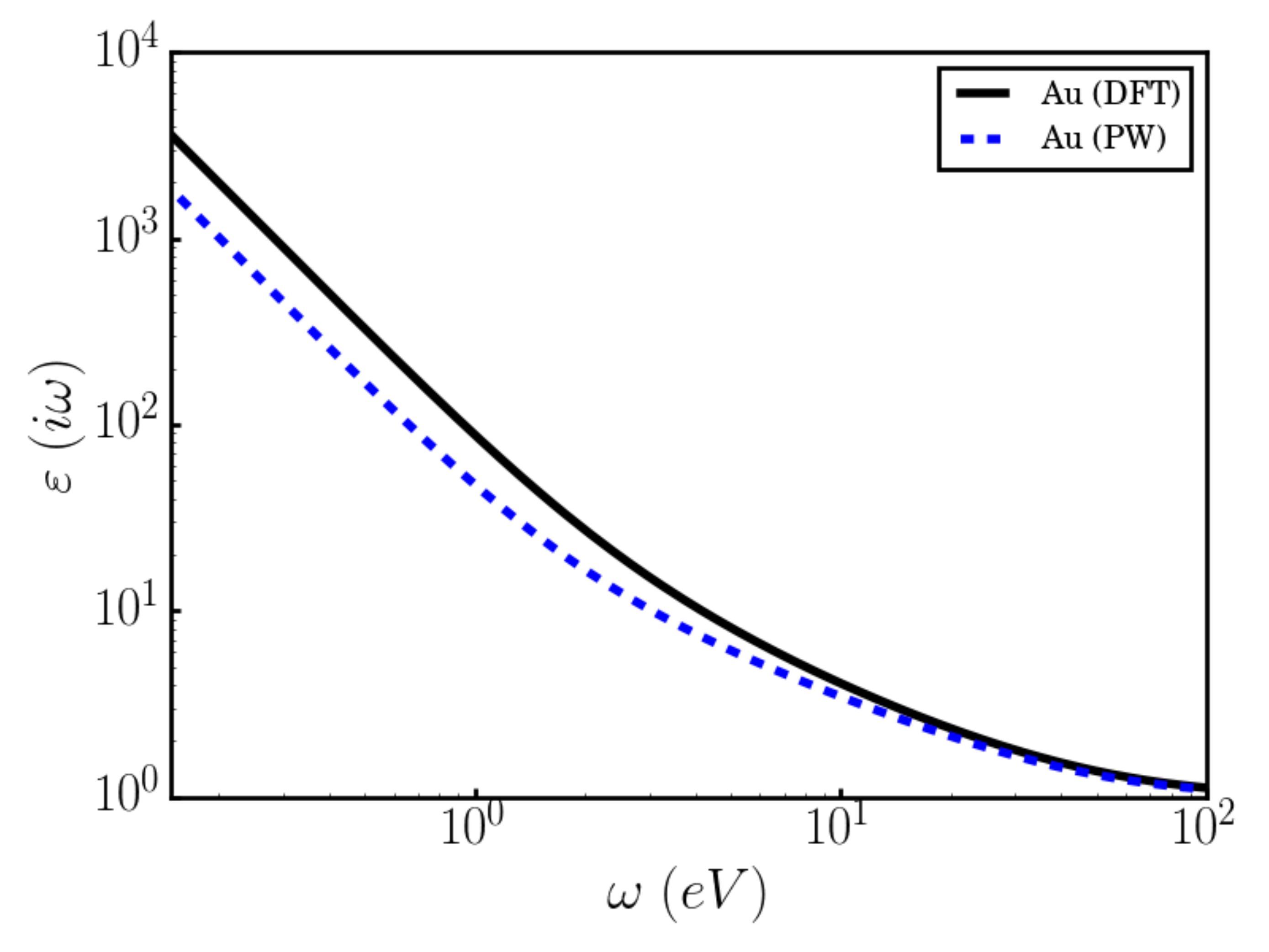}
\caption{ (Color online) The dielectric function at imaginary frequencies for gold from DFT calculations (solid curve).  The static value, $\epsilon(0)$ is 2.0$\times10^{5}.$ The exact value for the very large zero frequency term in the dielectric function does not substantially alter the final results in the present work. We also show an alternative dielectric function for gold (dashed curve) given by Parsegian and Weiss.\,\cite{PW}}
\label{figu2b}
\end{figure}

To calculate the Lifshitz
forces, we follow Lifshitz and co-workers' work\,\cite{Lif},
which involves expressing the Lifshitz interaction energy
per unit area ($E$) in terms of the electromagnetic
normal modes. The details for calculations
of this kind of geometries were outlined recently by Dou et al..\,\cite{Ser,Dou}
The Lifshitz pressure is given by the negative of  the
derivative of the Lifshitz energy with respect to the water
film thickness.
In the present work, we  calculate the Lifshitz interaction acting on premelted water sheet on ice inside a large pore  filled with a dilute gas.
In terms of the model geometry, we study a coated planar object interacting with a second planar object  in a medium, i.e., for the geometry $1|2|3|4$ (see Fig. 1).
Medium  1 is the surface material ($\epsilon_1$), 2 is dilute gas   ($\epsilon_2\approx1$), 3 is water ($\epsilon_3$), and 4 is the ice ($\epsilon_4$). 
To predict dielectric function of different materials, we carried out first principles calculations within the
density functional theory (DFT) 
 using the Vienna Ab Initio Simulation Package (VASP). For all systems, the exchange-correlation term is treated using the Perdew-Burke-Ernzerhof (PBE)~\cite{PBE} functional. Since PBE is known to underestimate the band gap energy, for both silica and alumina (corundum structure), we used scissors-operator correction to reproduce the experimental band gap energies ($\Delta=2.9$ for alumina  and $\Delta=3.6$ eV for silica). The Brillouin-zone integrations are performed using the $\Gamma$-centered Monkhorst-Pack\,\cite{Monkhorst} grids of $80\times80\times80$ and $18\times18\times9$ for Au and Al$_2$O$_3$, respectively. For mica we use a model dielectric function presented in the literature.\,\cite{Chan} For calculations of Au  and SiO$_2$ based systems, the details are given in our previous work.\,\cite{SashaPCCP2016,MathiasGoldEPJB} We test how sensitive the results are to the details of the dielectric functions by also using an alternative dielectric function for gold given by Parsegian and Weiss.\,\cite{PW} Unlike the temperature dependence of the Casimir force between metal surfaces\,\cite{Bost2000,Ser2001,Lamo,Hoye,Sush,Milt2,Decca} the exact values for the low frequency dielectric function of gold are not very  important in the present systems. It is enough to know that it is very large compared to the other dielectric functions for low frequencies. The ionic contribution to the static dielectric constant was determined from the
ion-clamped dielectric matrix by means of perturbation method.~\cite{WuVanderbiltHamann}
The dielectric functions of ice and water   (at $T=273.16$ $K$) were taken from the work of Elbaum and Schick.\,\cite{Elbaum}    The dielectric functions used are shown  in Figs.\,\ref{figu2a} and \ref{figu2b}. We observe that the dielectric function of water\,\cite{Elbaum}, ice\,\cite{Elbaum}, silica\,\cite{SashaPCCP2016} and gold\,\cite{PW} can be found as parameterised dielectric functions in the literature. Together with the Lifshitz theory in multilayer systems, described in many papers ( {see for instance previous works\,\cite{Ser,Lif,Dou, EstesoJCP, Tomas, Zhao}}), this makes our main results straightforward to reproduce.
 The  dielectric functions for ice and water have in the past also been used in a study as approximate dielectric functions for water
in different phases, at other temperatures and pressures, in a
study of van der Waals (vdW) interactions involving methane gas
hydrates (i.e. a mixture of methane and ice/water
molecules).\,\cite{Bonnefoy}   
The minerals that make up rock also depend on how the rock was formed. It is common that rock contains large amount of silica as  a large fraction of Earth crust is made up of quartz.  
The kind of rock or soil materials that constitutes the pore surface influences the premelting, i.e. the formation of a water film on the ice surface.
To quantify how sensitive premelting is to the  material, we base our investigation on several different earth minerals. In particular, we study
different silica materials (one resembling quartz), as well as alumina, and  gold.

\begin{figure}
\includegraphics[width=8cm]{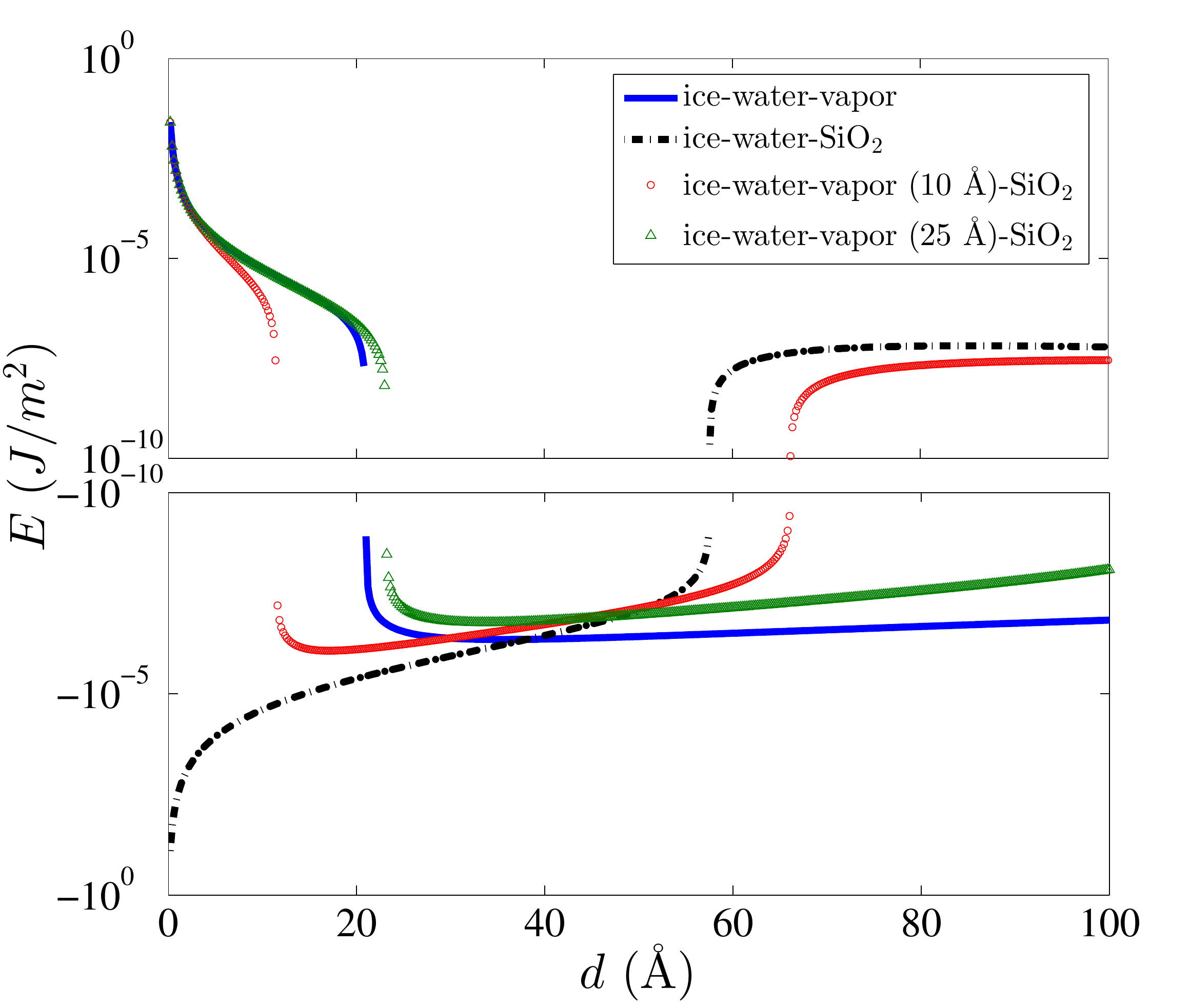}
\caption{(Color online)  Lifshitz interaction energy per unit area, $E$, (across waterfilm with width $d$) for ice-water-vapor, ice-water-quartz, ice-water-vapor-quartz (10 \AA \ vapor thickness), and ice-water-vapor-quartz (25 \AA\ vapor thickness) systems.  }
\label{figu3}
\end{figure}

\begin{figure}
\includegraphics[width=8cm]{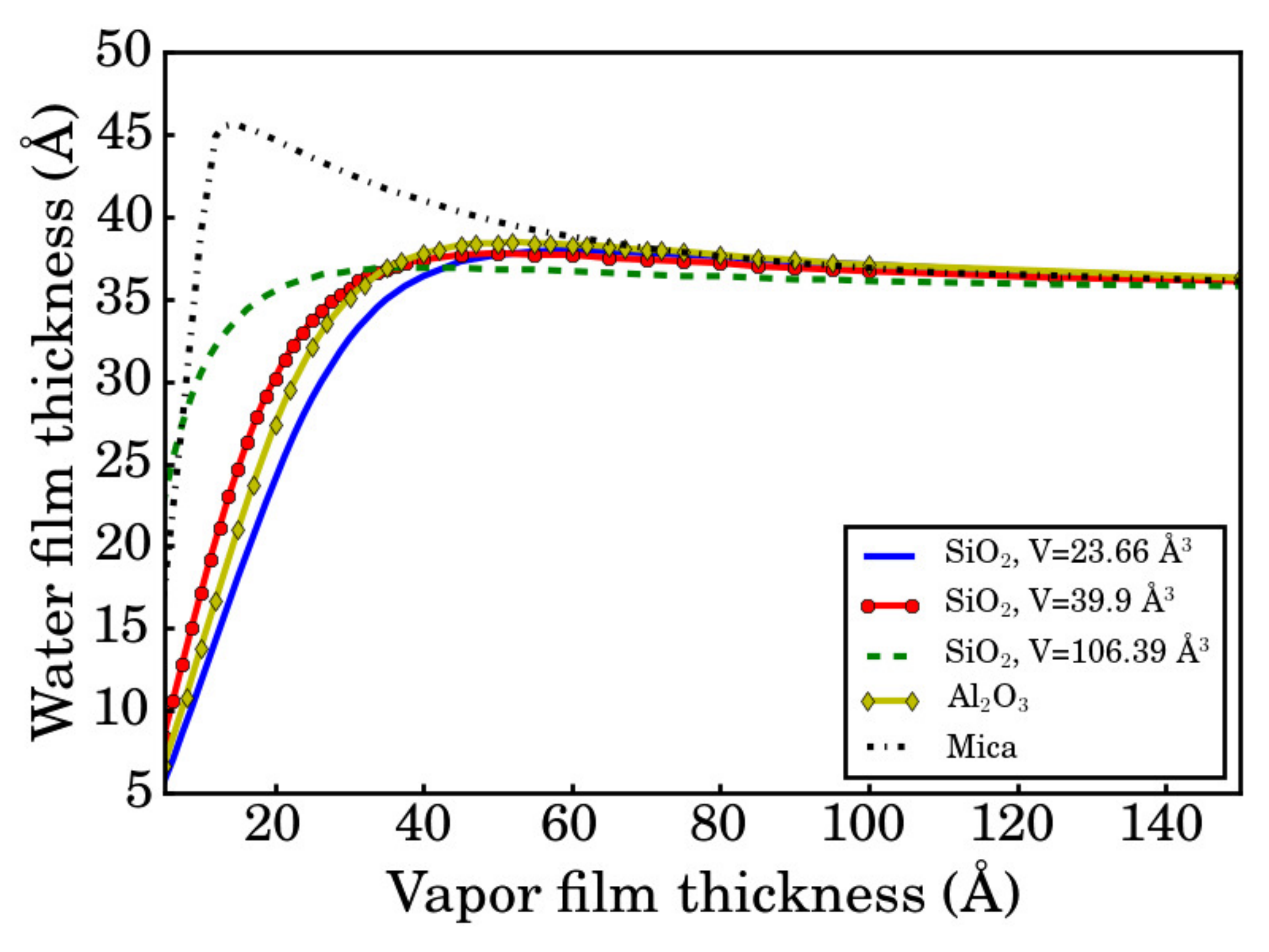}
\caption{(Color online) Equilibrium water film thickness versus gas layer thickness for different pore surfaces: (1) SiO$_2$ (V=23.66  \AA$^3$ ); (2) quartz (SiO$_2$ with V=39.9 \AA$^3$); (3) SiO$_2$  (V=106.39 \AA$^3$);  (4) Al$_2$O$_3$; (5) Mica\,\cite{Chan}.  Details for silica are given in Ref. \,\cite{SashaPCCP2016}.}
\label{figu4}
\end{figure}

Figure \,\ref{figu3} shows calculated Lifshitz interaction energies across thin water films of different thicknesses. As the minimum of the energy curves indicates the optimal waterfilm width, the figure shows that  a thin vapor sheet between ice and pore surface is essential  for premelting to occur. 
For the case of very thin gas layer between the ice and the pore surface, the Lifshitz forces lead to reduced sheets of liquid water. 
In between thick vapor layers and the case with no gas layers, the figure shows that in a certain thickness range for the gas region, the
water film thickness can be enhanced. 
In Fig.\,\ref{figu4},
we have also plotted, for  
different silica materials, alumina, and mica  the water film thickness as a function of vapor thickness.
For the case of a silica surface (with properties similar to quartz), we find around 6-7 $\%$ widening of the water film thickness
for a vapor film 5~nm wide within a pore compared to in the regular  ice-water-vapor system.  This result is related to the fact that a system with ice-water-silica has both short range attraction and long range repulsion.  Similar results are observed for alumina where an 8$\%$ widening  can be seen.
 The widening   can reach up to almost 30\% for mica.  Nanoscale curvature of ultrasmall pores  in silica or alumina may introduce additional effects on film thickness but less so for the planar
mica material.
These materials are common  in rock and soil and appear to behave in a similar way. Frozen soil in general is typically a mixture of quartz, clay, and organic
material (humus). 
The quartz (in rock or soil) can weather, making it porous. 
Molecules (e.g. organic or water) can then enter pores to
coat the surface. 
Another important rock/soil material  is clay. 
There are a handful of different clays  such as mica.
The common feature is that they  are formed of aluminosilicate layers.   
  Different clays have different
sequences of silica and alumina forming the single aluminosilicate
layer.  Between the aluminosilicate layers are "air gaps" called
galleries.  When the ice filled clay becomes wet at the ice interface, water could penetrate into the galleries,
then the gap of the gallery gets wider and the clay swells.

\begin{figure}
\includegraphics[width=8cm]{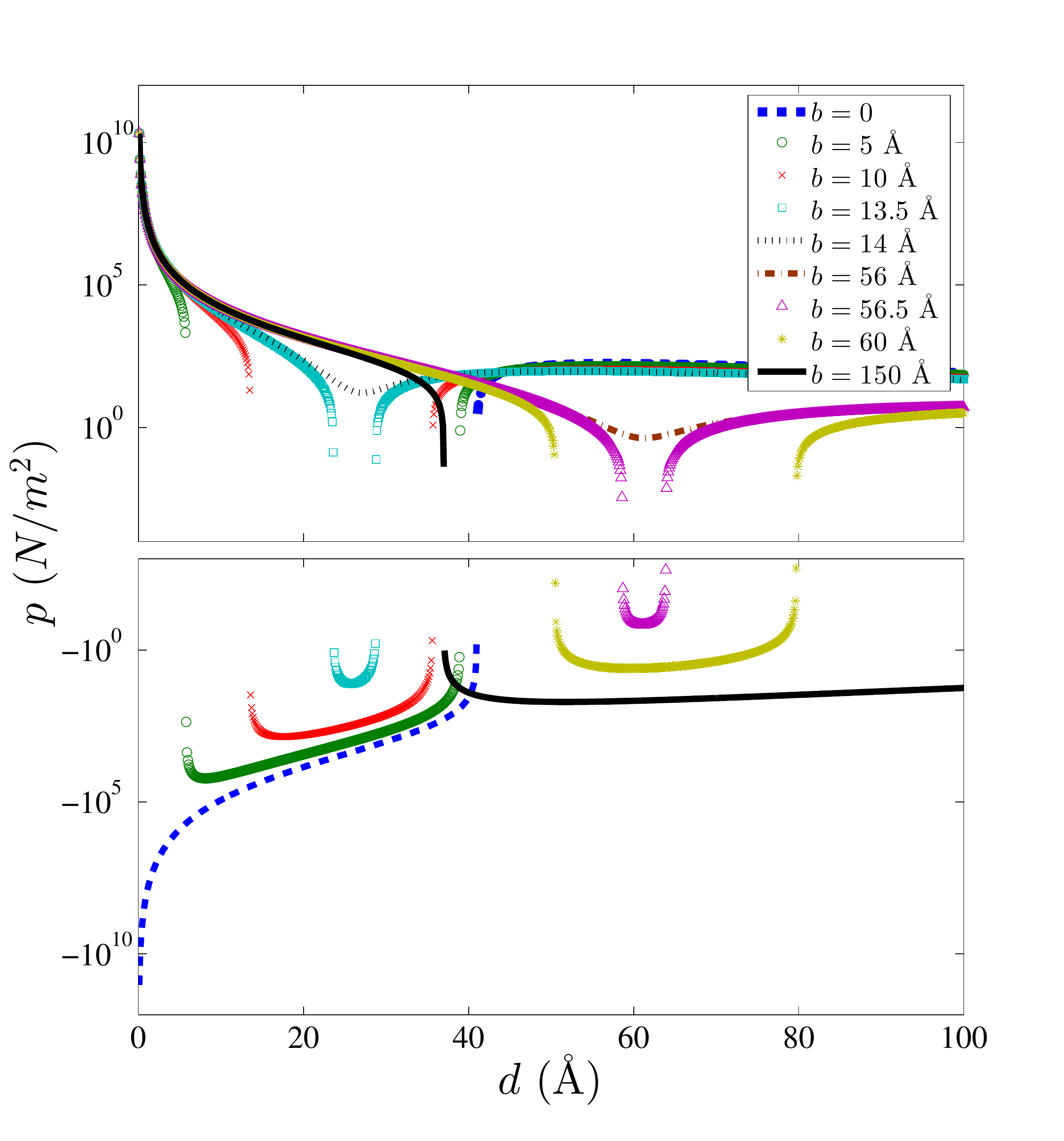}
\caption{(Color online)  Lifshitz pressure, $p$, across waterfilm with width $d$ for ice-water-vapor-gold system. The thickness of the vapor layer ($b$) is given in the caption. For the thickest vapor layer shown the repulsion-attraction transition goes down towards the limiting value for ice-water-vapor ($\rightarrow$ 35.5 \AA). Positive pressure across the water film corresponds to repulsion. In the vapor thickness range between roughly {14 \AA} and {56 \AA} the pressure is positive for all water film thicknesses indicating complete surface melting.}
\label{figu5}
\end{figure}

\begin{figure}
\includegraphics[width=8cm]{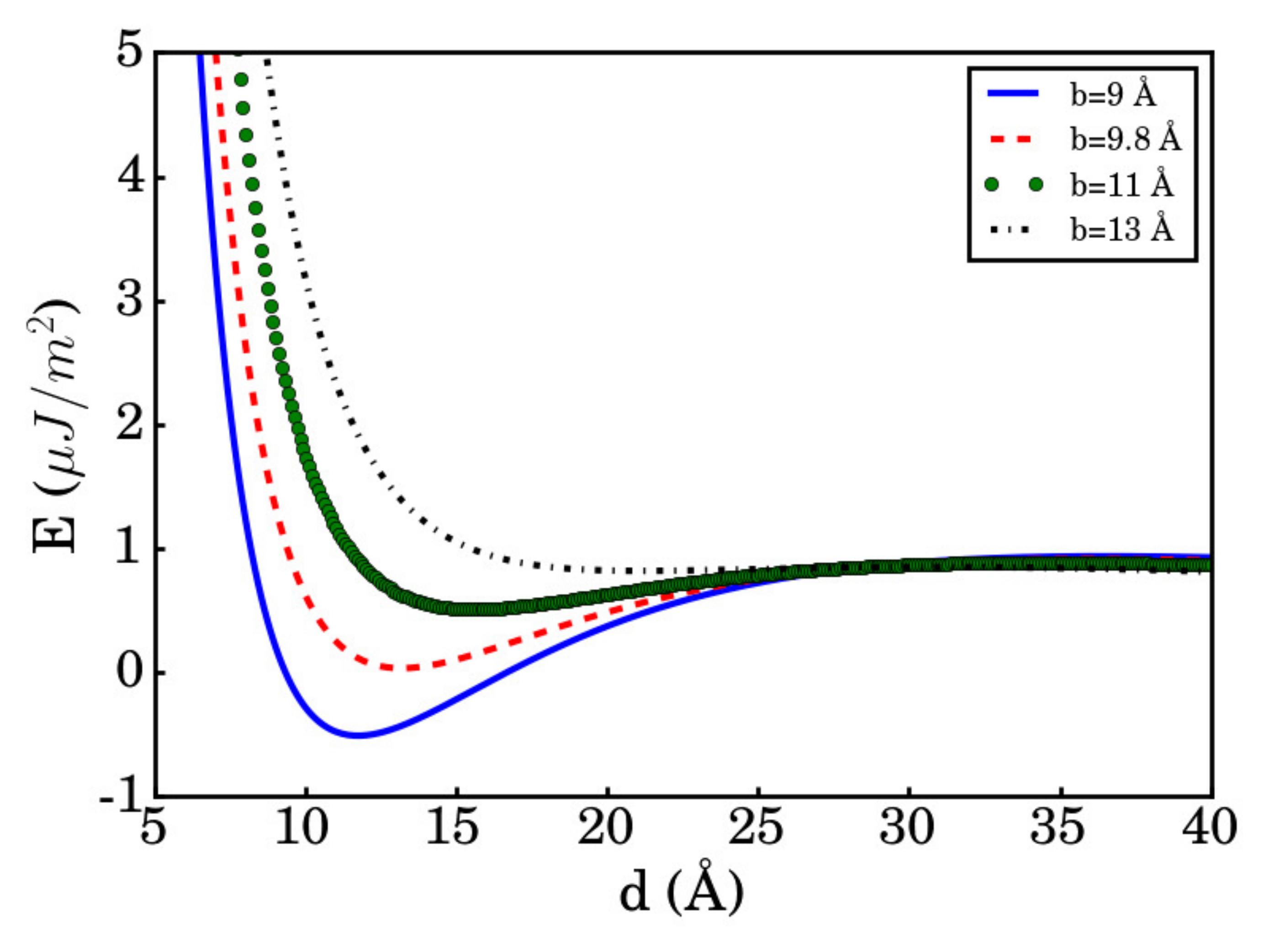}
\caption{(Color online)  Lifshitz interaction energy, $E$, across waterfilm with width $d$ for ice-water-vapor-gold system. The thickness of the vapor layer ($b$) is given in the caption. A detailed study for b values near the transition between region II and region III  ($b\approx$~9.8 \AA).  }
\label{figu7}
\end{figure}

\begin{figure}
\includegraphics[width=8cm]{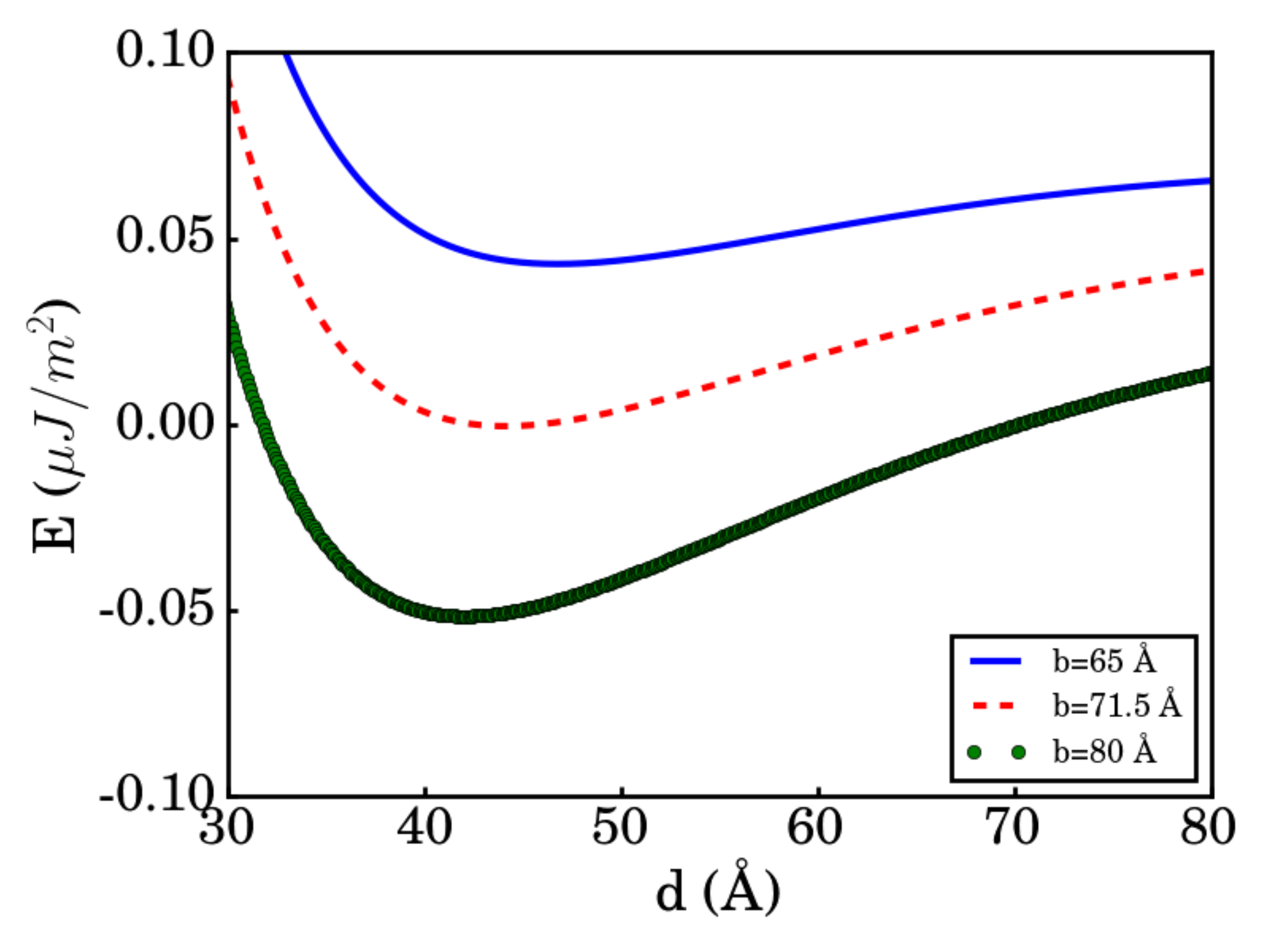}
\caption{(Color online)  Lifshitz interaction energy, $E$, across waterfilm with width $d$ for ice-water-vapor-gold system. The thickness of the vapor layer ($b$) is given in the caption. A detailed study for b values near the transition between region IV and region V ($b\approx$~71.5 \AA). Note that one needs to go down to the transition thickness ($b\approx$~56 \AA) to have a curve without a local energy minima.  }
\label{figu8}
\end{figure}

Lifshitz pressures for ice near a metallic surface patch are shown in Fig.\,\ref{figu5}. Here we used dielectric function for gold from DFT calculations. At positive pressures, the water film thickness increases, while at negative, the thickness decreases.
In the absence of a vapor layer, the ice surface stays dry as there is no short range repulsion.  The Matsubara terms in the Lifshitz interaction energy contribute in an attractive manner when gold is near the water and ice but in a repulsive manner for large separations.  
This repulsive contribution becomes larger than the attraction for thicker vapor layers.
The water film width increases with increasing vapor film thickness until the Lifshitz forces suddenly give rise to complete surface melting, as indicated by the positive pressure for all film thicknesses.
As the vapor film increases further, the liquid film thickness becomes finite once more and decreases down towards that of ice-water-vapor limit with increasing vapor thickness. These two discontinuous transition points occur at vapor film thicknesses close to {14 \AA} and {56 \AA} (or close to {10.5 \AA} and {55 \AA} when  using an alternative dielectric function for gold\,\cite{PW}).
Complete surface melting occurs between the two  points. A similar effect was observed for ice-water-tetradecane-vapor where a single transition point occured.\,\cite{Bar-Ziv} 
The origin of the first transition point in our system is that
for a range of frequencies, gold  is more polarizable than the
water which is more polarizable than ice, thus
enhancing those components of the vdW interactions
which promote thickening of the water film (cf. Ref. \,\cite{Bar-Ziv}).
The extreme sensitivity with respect to small separations  follows from  the mathematically  complex Lifshitz formalism and can hardly be given a simple physical interpretation. It is in this context of interest to recall how a vdW attraction can surprisingly enough  be reversed into a Casimir repulsion even at very small distances because of  retardation  effects  (cf. Ref. \cite{bosserPRA2012}).
We find that above the second transition point, the film decreases  towards {35.5 \AA} which is the limiting value for a three layer system with  ice-water-vapor as predicted by Elbaum and Schick.\,\cite{Elbaum}.  
The above considerations limit the region of vapor film thicknesses where there will be complete melting (with global minima at infinity and without any local energy minima where water films could be trapped at a finite film thickness). To gain
further insight into this, following Bar-Ziv and Safran\,\cite{Bar-Ziv}, we present in Figs.\,\ref{figu7} and \,\ref{figu8} the Lifshitz  interaction energy versus water film thickness for the system with ice-water-vapor-gold (using same dielectric functions as in  Fig.\,\ref{figu5}). Here we can clearly see that there are two additional regions of interest. 
The energy study reveals that there are all together 5 vapor thickness regions of interest: (I) {$b$~=~0-9.7 \AA} leads to finite size water films; (II) {$b$~=~9.8-13.5 \AA} can lead to complete melting but it can also be  inhibited due to a local energy minima to form finite water film; (III) {$b$~=~14-56 \AA} leads to the case above with complete melting; (IV)  {$b$~=~56-71 \AA} has the potential for complete melting that can be inhibited due to a local energy minima to form finite water film; (V) {$b\geq$~71.5 \AA} leads to a finite water film that approaches that for ice-water-vapor system for large $b$ values. The water film thickness is around 4 nm just above the point between region IV and V.
In region II and IV thermal fluctuations are highly likely to induce the system into complete surface melting. The exact values of the transition points vary with the details of the dielectric functions used. In real systems they also depend, for instance, on potential presence of salt ions.
 In the 4-phase system modelled here, the solid  phase is exposed only to vapor and has no direct contact with the water film. Hence the surface is assumed to be uncharged. When in contact with water, silica and alumina will develop a surface charge as acidic surface protons dissociate in solution \cite{Parsons2009}. At the same time, salt lowers the melting point of water, an effect that can be enhanced under confinement in pores \cite{meissner}. In this case the surface charge together with dissolved salt  may generate a water film thicker by  orders of magnitude \cite{Wettlaufer}.  {In the presence of salt, the $n=0$ Matsubara term becomes screened by the ions.\,\cite{Maha, YaminskyNinham, DaviesNinhamRichmond} However, this is expected to play a minor role.}

We conclude that liquid water can exist in ice filled pores and cavities.  As we have discussed surface premelting occurs when a thin film of water forms on the ice surface caused by repulsive Lifshitz forces. 
 For the thin film widths considered here, gravitational force can be neglected but ion specific dispersion forces between any ions present and the surfaces could influence the result.\,\cite{BWN2001,Thiyam} 
We suggest that the same mechanism may be found on the surface of permafrost regions. 
We predict that the presence of water sheets (created by surface premelting) on ice in pores depends on the presence of a gas layer between the ice and surface.   In the studied systems, we find that no premelting layer can exist when an ice surface is too close to the pore surface. 
Premelting in porous media also depends on the rock or clay (soil) composition. Under specific conditions, it is even possible to  have very slightly enhanced  liquidlike water content for ice filled pores in quartz rock (modelled as silica) compared to when ice is in free space. 
The results with pore surfaces made up of silica and alumina are similar. There is a bigger enhancement effect for pores with mica surfaces.
 In pore regions with metallic surface patches, there can in the presence of thin vapor layers be complete melting.
Having two transition points, between a system with finite size water films and a system with complete surface melting (and back again to a system with finite water film), is a new prediction based on Lifshitz forces in multilayer system. The exact values for these transition points depend on the detailed experimental dielectric functions of actual materials.
Strictly speaking there are more transition points between systems with (a) region I (V) where there is a finite size film and (b) region II (IV) where there is complete melting that can be inhibited due to a local energy minima to form finite water film.
 Such dramatic effects from surface materials and gas layers on melting of ice in pores have not been accounted for in estimates for the amounts of liquid water present in the subsurface of icy planets.
Our results may provide additional insight into the conditions under which liquid water could be present in ice filled pores and cavities.   Liquid water plays a crucial role in life processes and the presence of microbes in the permafrost and in the Greenland glacial ice has been linked to liquid water forming along grain boundaries in polycrystalline ice and on mineral surfaces such as clay grains.\,\cite{Price-cold,Tung1}
  In fact, confined water regions play a central role in theories concerning the cold origin of life \cite{Price-cold}.  
Given the tiny dimension involved and the role of the dielectric function of the material, 
our work could stimulate efforts to include the dielectric function of the microbe itself to better understand how it can access water in such habitats. 
The predicted premelting of ice in the presence of gold grains and thin vapor layers indicates that metals could bring about water-rich patches within the permafrost, possibly providing habitats for life.

We acknowledge support from the Research Council of Norway (Projects 221469 and 250346).  
 We acknowledge access to high-performance computing resources via SNIC and NOTUR.   PT acknowledges support from the European Commission; this publication
reflects the views only of the authors, and the Commission cannot be
held responsible for any use which may be made of the information
contained therein. Useful discussions with  Dr K.V. Shajesh are gratefully acknowledged.


\end{document}